\documentclass[fleqn,twoside]{article}
\usepackage{espcrc2}
\usepackage{graphicx}
\usepackage[figuresright]{rotating}

\newcommand{\AmS}{{\protect\the\textfont2
  A\kern-.1667em\lower.5ex\hbox{M}\kern-.125emS}}
\newcommand{\be}{\begin{eqnarray}}
\newcommand{\ee}{\end{eqnarray}}

\newcommand{\ra}{\rangle}

\title{Studying the role of the 't~Hooft %instanton-induced 
interaction in QCD, by means of lattice simulations
}
\author{P. Faccioli\footnote{Speaker}\address{E.C.T.*, 
Strada delle Tabarelle 286, I-3905 Villazzano (Trento), Italy.},
T.A. DeGrand\address{University of  Colorado, Boulder, CO 80309, USA.}}
\begin{document}

% typeset front matter

\begin{abstract}
We report on a recent investigation on 
the dynamics of light quarks, at intermediate and low-energy. 
By measuring  some appropriate correlation functions on the lattice, 
it is possible to probe the Dirac and flavor structure of the non-perturbative 
quark-quark interaction and look directly for signatures
of the 't~Hooft Lagrangian. Results obtained with chiral fermions
strongly support the instanton picture and the phenomenological parameters
of the Instanton Liquid Model.
\end{abstract}
\maketitle
\section{Introduction}

%Understanding the structure of the quark-quark interaction 
%at all scales requires the solution of the non-perturbative sector of QCD.
%Although this program has not yet been carried out completely, 
%important information has been gathered in the last two decades.
%In particular, it has become clear that the microscopic non-perturbative
%dynamics which is relevant for the physics of light hadrons 
%is quite different from that underlying heavy hadrons. This is because
%light quarks are much more sensitive to the gauge configurations responsible 
%for the spontaneous breaking of chiral symmetry.

Lattice QCD simulations have played a major role in improving our 
understanding of the quark dynamics.
For example, by measuring the Wilson loop it was possible to extract 
the non-relativistic confining potential, between two static color sources.
Unfortunately, the concept of a non-relativistic 
potential becomes useless, when
the color sources involved are light or even massless dynamical quarks.
Clearly, any information about the interaction between $u$ and $d$ quarks
must be cast into a fully relativistic field-theoretic formalism.
This makes the problem of understanding the 
light-quark dynamics much more complicated. 

An important step forward in this direction was made by 
't~Hooft \cite{'tHooftU1},
who derived a low-energy effective contact interaction between light quarks, 
by computing  the Euclidean QCD partition function %generating functional 
in the semi-classical limit 
(i.e. accounting for small perturbations around the instanton solution).
In the $N_f=2$ case, the 't~Hooft Lagrangian can be written as:
\be
\label{Lthooft}
L_{tH}=
C_{\bar{\rho},\bar{n}}\,(\,\frac{2\,N_c-1}{2\,N_c}\,
[\,(\psi^\dagger\,\tau^-_a\,\psi)^2%$\\
\nonumber\\
- (\psi^\dagger\,i\gamma_5\,\tau^-\,\psi)^2\, ]
+\frac{1}{4\,N_c}\,(\psi^\dagger
\,\sigma_{\mu\,\nu}\,\tau_a^-\,\psi)\,),%$
\ee 
where $\tau^{-}=(\vec{\tau},i)$ ($\vec{\tau}$  are isospin Pauli matrices), and
$C_{\bar{n}\,\bar{\rho}}$ is a coupling constant depending on
the typical density of instantons in the vacuum $\bar{n}$,
and  their typical size $\bar{\rho}$.
%:
%$ L_{V-AV}= C' \,\left[
%\,(\psi^\dagger\,\gamma_\mu\,\tau^-_a\,\psi)^2 
%+ (\psi^\dagger\,\gamma_\mu\,\gamma_5\,\tau^-\,\psi)^2\, \right]$.
%
%Let us examine the specific features of the  
%'t~Hooft Lagrangian:
%\footnote{Obviously, L_{\tH} is 
%not the most general lowest-dimensional 
%contact interaction compatible with the QCD symmetries.
%For example, it is possible to add a combination of
%vector (V) and axial-vector (AV) terms 
%and still respect the requirement of invariance under chiral 
%rotations. These terms are indeed included in 
% the Nambu Jona-Lasinio model \cite{NJL}} L_{\tH}.

The most important feature of the 't~Hooft interaction is
 that it is invariant under global
 $SU_L(2)\times SU_R(2)$ rotations, but changes sign under $U_A(1)$ 
transformations. Moreover, the vertex is active only between quarks of
 \emph{different flavor and different chirality}. 
%This means that a the 't~Hooft interaction contributes at Leading Order (LO)
%to the pseudo-scalar meson wave-function, but only at Next-to-Leading 
%Order (NLO) to the vector meson wave-function.
%Notice also that, due to the presence of S and PS coupling,
%the chirality of each quark is flipped after each single interaction.
%Finally, the vertex is active only 
%between quarks of different flavors. 
%This means that a 't Hooft interaction contribute at LO to the baryon octet 
%wave-functions but only at NLO to the baryon decuplet wave-function. 
%\end{itemize}
The finite size of the instanton field
provides a natural cut-off scale for the interaction, 
$\Lambda_{cut-off}\simeq~1/\bar{\rho}$.
% and a prescription on the cut-off 
%procedure\footnote{High momentum modes should be removed by delocalizing the
%contact interaction}.

Unfortunately, the  instanton density and size distribution
cannot be computed from first principles in QCD.
%, therefore
%we do not know the strength  of the 't~Hooft interaction and the momentum 
%scale above which it becomes negligible.
However, the Dirac and flavor structure of the 't~Hooft Lagrangian 
is a model-independent prediction of the semi-classical approximation.
The Instanton Liquid Model\footnote{
In the last two decades it has been shown that the ILM can quantitatively 
explain important non-perturbative phenomena related to the structure of the  
QCD  vacuum, as well as  light hadrons masses
and formfactors (see e.g. \cite{shuryakrev,masses,formfactors}}
  (ILM) \cite{shuryakrev}
consists of assuming that the vacuum is saturated
by instantons with typical size $\bar{\rho}\simeq 1/3$~fm and density 
$\bar{n}\simeq 1\,\textrm{fm}^{-4}$.

%A number of approaches have been proposed so far in order 
%to check the instanton picture of the QCD vacuum
%by means of lattice  QCD simulations (see e.g. \cite{cooling}
% and \cite{zmpeaks}).
%The cooling procedure  consists on 
%studying the behavior of the local topological charge, 
%when one averages over ensembles of configurations 
%which are closer and closer to the extreme of the action. 
%As quantum fluctuations are progressively suppressed,
%topology is seen to cluster around smooth ``lumps'' of gauge, 
%which can be identified with instantons.
%An alternative approach is based on the observation
%that instantons generate a symmetric distribution of the chirality of 
%the low-virtuality modes of the Dirac operator \cite{zmpeaks}, picked 
%around +1 and~-1. 

In the following, we shall report on a recent lattice-based investigation
\cite{chimix,scalar}, which  allows to check if 
effective quark-quark interaction in QCD has the specific Dirac and flavor 
structure predicted by the semi-classical 
expression $L_{tH}$, and if the ILM phenomenological parameters
are realistic. 

%We shall first define some correlators which, on the one hand, 
%allow to probe directly the Dirac and flavor structure of L_{\tH} 
%and, on the other hand, receive no perturbative contribution.
%Then, we shall match instanton-based predictions for such correlators
%against lattice simulations.
%The result our analysis strongly support the instanton picture and the ILM.

%The paper is organized as follows.
%In section \ref{Dirac} we focus on the Dirac structure
%and compare lattice result with ILM simulations. 
%In section \ref{flavor} we show how the
%analysis can be extended to probe the flavor structure of L_{\tH}.
%In section (\ref{loops}) we shall study how the effective
%interaction changes, when quark loop are suppressed. 
%Results and conclusions are summarized in section \ref{conclu}.
\section{Dirac structure}
\label{Dirac}

In order to probe the Dirac structure of $L_{tH}$,
we  study  
the flavor Non-Singlet (NS) chirality-flip ratio,  introduced in \cite{chimix}:
\be
\label{RNS}
R^{NS}(\tau):=\frac{A^{NS}_{flip}(\tau)}{A^{NS}_{non-flip}(\tau)}=
\frac{\Pi_\pi(\tau)-\Pi_\delta(\tau)}
{\Pi_\pi(\tau)+\Pi_\delta(\tau)},
\ee
where $\Pi_\pi(\tau)$ and $\Pi_\delta(\tau)$ are pseudoscalar (PS) and 
scalar (S) iso-triplet
two-point correlators related to the currents $J_\pi(\tau):=\bar{u}(\tau)
\,i\gamma_5\,d(\tau)$  and $J_\delta(\tau):=\bar{u}(\tau)\,d(\tau)$.
If the propagation is chosen along the  
time direction, $A^{NS}_{flip(non-flip)}(\tau)$ represents 
the probability amplitude for a 
$|q\,\bar{q}\ra$ pair with isospin 1 to be found after a 
``time'' interval $\tau$ in a state
in which the chirality of the quark and anti-quark 
is interchanged (not interchanged).
Notice that the ratio $R^{NS}(\tau)$ 
must vanish as $\tau\to 0$ (no chirality flips),
and must approach 1 as  
$\tau\to \infty$ (infinitely many chirality flips).
It can be shown that the correlator (\ref{RNS}) 
receives no leading perturbative contribution.
% and
%probes directly the S and PS components of the effective vertex.
%A model-independent spectral analysis of $R^{NS}(\tau)$
%indicated that the interaction responsible for dynamical chirality mixing
%is mediated by topological fields and that
%the rate of chirality flips in a quark-antiquark system
%is given by the mass of the $\eta'$ meson.

\begin{figure}
\includegraphics[scale=0.30,clip=]{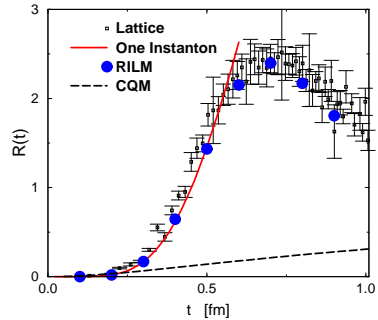}
\caption{Results for $R^{NS}(\tau)$. 
Squares are lattice points of \cite{degrand1}, stars are RILM points.
The significance of the other curves is
discussed in \cite{scalar}.}
\label{mainfig}
\end{figure}

The two-point functions
appearing in (\ref{RNS}) have been 
%first calculated in the
%quenched approximation by the MIT group \cite{lattice2pt} with 
%Wilson fermions and more 
recently  calculated by one of the authors,
using chiral (overlap) fermions \cite{degrand1}.
Different results for $R^{NS}(\tau)$ 
are presented in Fig.~\ref{mainfig}.
%As expected, the lattice data interpolate between 0 and 1. 
Notice that the lattice curve (square points)
% has a maximum
%at about 0.7~fm, where the ratio is 
becomes considerably larger than one.
This implies that, after few fractions of a fermi, 
the quarks are most likely to be found
in a configuration in which their chiralities is flipped.
%, than to be
%found in their initial configuration.

Let us now  consider the result for  $R^{NS}(\tau)$ obtained 
in the Random Instanton Liquid Model (circles), which
accounts for the 't~Hooft interaction to
all orders, but neglects quark loops (quenched approximation).
The agreement with the lattice results is 
impressive.
The presence of a maximum in $R^{NS}(\tau)$
and the subsequent fall-off toward 1 
have a very simple explanation in the ILM: 
if quarks propagate in the vacuum for a time comparable with
the typical distance between 
two neighbor instantons (i.e. two consecutive 't Hooft interactions), 
they have a large probability of crossing the field of the closest
pseudo-particle. 
If so happens, they must necessarily flip their chirality, due to the 
S and PS structure of the 't Hooft vertex.
So, after some time, the quarks are most likely to be found in the
configuration in which their chirality is flipped.
On the other hand, 
if one waits for a time much longer than 1~fm, then the quarks will ``bump''
 into many other such pseudo-particles,
experiencing several more chirality flips. Eventually, either chirality 
configurations will become equally probable 
and $R^{NS}(\tau)$ will approach 1.

%The position of the maximum in $R^{NS}(\tau)$ carries
%information about the interplay between one-body
%and many-body effects generating chirality mixing (see the detailed
%discussion in \cite{scalar}).
%It is interesting to compare the above numerical
%RILM results with the single-instanton contribution (solid
%line in Fig. \ref{mainfig}), which
%was derived analytically by one of the authors in \cite{chimix}.
%From such a comparison, one can see that one-body effects,
%i.e. chirality flips induced at the level of a {\it single} interaction, 
%dominate the ratio up to Euclidean times of the order of 0.5-0.7 fm. 
%The onset of many-body (many-instanton) effects is 
%governed by the numerical value of the instanton density: 
%the less dilute is the system, 
%the earlier many-instanton effects become important. 
From the agreement between ILM and lattice data one may argue that
the phenomenological 
values $\bar{\rho}\simeq 1/3$~fm, 
$\bar{n}\simeq~1\,\textrm{fm}^{-4}$ are indeed realistic.

%The prediction for $R^{NS}(\tau)$ would be
%drastically different if the leading non-perturbative interaction
%was predominantly chirality conserving, i.e. if the
%effective Lagrangian  contained only V or AV coupling.
%In this case, the chirality mixing would  only be induced by the dynamical
%mass generation  due to the spontaneous breaking of chiral symmetry 
%(i.e. by a genuine many-body effect). 
%In fact, unlike in the ILM,
%a  {\it single} quark-antiquark V and AV
%interaction would not interchange the chirality of quarks.
%As a result, even in the quenched approximation, 
%quarks would {\it never} more likely to be found in the 
%flipped chirality state than in the initial chirality state, i.e. 
%$R^{NS}(\tau)<1$ for all $\tau$.
%On a {\it qualitative} level, the prediction for $R^{NS}(\tau)$
%of any model with a purely V or AV coupling 
% would be similar to that
%obtained considering the propagation
%of a free but {\it massive} ``constituent'' quark and anti-quark pair in the
%vacuum (dashed line in Fig.~\ref{mainfig}).
%From this analysis we  can conclude
%that the presence of a maximum in the lattice results for
%$R^{NS}(\tau)$ implies that  the non-perturbative
%quark-quark interaction contains a strong S and PS component.
%In other words, the chirality is mixed already at the level
%of a {\it single} quark-quark interaction, and not only through 
%collective many-body effects.
%Hence, these lattice results rule-out scenarios in which the 
%interaction is predominantly V and AV.

\section{Flavor structure}
\label{flavor}

In this section, we shall introduce another correlator, which can 
be used to infer simultaneously information about the {\it chiral and 
flavor} structure of the effective vertex.
We define:
\be
\label{Rflavor}
R^{tH}(\tau):= \frac{(\Pi_\pi(\tau)-\Pi_\delta(\tau)) + (\Pi_\sigma(\tau)
-\Pi_{\eta'})}{(\Pi_\pi(\tau)-\Pi_\delta(\tau)) - (\Pi_\sigma(\tau)
-\Pi_{\eta'})}.
\ee
On the one hand, the numerator receives no perturbative contribution, but
couples maximally to the 't~Hooft interaction. 
On the other hand, the denominator does not  
receive perturbative contribution, and does not receive 
contribution from the single-instanton 't~Hooft interaction, as well. 
Hence, if instantons were the only sources of non-perturbative 
interactions in the light-quark sector of QCD, 
then $R^{tH}(\tau)$ should display a dramatic enhancement (divergence), 
in the limit $\tau\to 0$ ( i.e. when the single-instanton
contribution becomes dominant over many-instanton effects).
On the contrary, if the driving non-perturbative interaction does not have
the specific chiral/flavor structure of $L_{tH}$, 
there should be no enhancement in the short distance limit, 
because the denominator in $R^{tH}(\tau)$ does not vanish.
As a result, the magnitude of $R^{tH}(\tau)$ near the origin
represents a model-independent
estimate of the strength  of the 't~Hooft 
instanton-induced interaction, relative to
other sources of non-perturbative dynamics. 
We suggest that this quantity
should be investigated on the lattice.

\section{Instanton, unitarity and quenching effects}
\label{loops}

We recall that lattice 
results reported in section \ref{Dirac} 
have been obtained in the quenched approximation. 
It is important to ask what differences should be expected in full QCD. 
%Using general QCD inequalities \cite{inequalityQCD}, 
It is immediate to show that $R^{NS}(\tau)>1$ if and only if 
 $\Pi_\delta(\tau)<0$.
The negativity of such a two-point function 
%represents a severe failure of the
%quenched approximation which appears only at sufficiently small values
%of the quark mass and  
is a reflection of the fact that the unitarity of the theory is lost.

In terms of chirality flipping amplitudes, 
we see that the $\Pi_\delta(\tau)\ge 0$ 
constraint implies that quarks must never be
more likely to  be found in the flipped chirality configuration 
than that in the original configuration. 
%$A_{flip}(\tau)\le A_{non-flip}(\tau)$.
Hence,  we can conclude
that the fermionic determinant suppresses some
chirality flipping events, which are otherwise allowed in the
quenched approximation.
Such a dramatic qualitative difference between quenched and full QCD
calculations of (\ref{RNS}) can be exploited in order to test any
 phenomenological description of the non-perturbative dynamics. 
{\it In fact, a realistic model must reproduce a dramatic enhancement of the 
chirality flipping amplitude, when quark loops are suppressed.}

Indeed, this phenomenon is naturally explained in the instanton picture
\cite{scalar}.  
Quark loops generate attraction between 
instantons and anti-instantons 
leading to a screening of the topological charge. %\cite{toposcreening}
As a result, quarks crossing the field of an instanton 
are very likely to find, in the immediate vicinity, an
anti-instanton which restores their initial chirality 
configuration.
%
%In Fig.~\ref{unquenching}(B) we compare the chirality flipping ratio
% $R^{NS}(\tau)$ obtained from a quenched (RILM)
%and unquenched (IILM\footnote{Interacting Instanton Liquid Model}) 
%version of the ILM.
With the inclusion 
of the fermionic determinant, 
the unitarity condition $R^{NS}(\tau)<1$ is actually restored. 
We  stress that, although such a restoration  
must necessarily take place in QCD, it represents a remarkable success 
of the ILM, which is not a unitary field theory.
%\begin{figure}
%\qquad\includegraphics[scale=0.45,clip=]{quenching2.eps}
%\hspace{1.5cm}
%\includegraphics[scale=0.35,clip=]{R_IV_ILM.eps}\\
%(A)\hspace*{7cm} (B)\\
%\caption{ (A):
%Suppression of chirality flips due to the topological
%screening induced by fermion-loops in the ILM ($N_f=3$).
%(B): Suppression of chirality flipping events, due to fermion-loop 
%exchange in the ILM. Circles are RILM (quenched) results, squares are IILM 
%(unquenched) results. In the unquenched model the unitarity requirement,
%$R^{NS}(\tau)\le 1$ is restored.}
%\label{unquenching}
%\end{figure}

\section{Conclusions}
\label{conclu}

We have presented a study of the light quark dynamics at low energy, based
on the results of lattice QCD simulations with chiral fermions.
Our analysis provides
an indication that the low-energy effective quark-quark
interaction has the Dirac structure predicted by the semi-classical
 the 't~Hooft interaction, and that the phenomenological 
parameters of the ILM are realistic.
We identified a correlator, $R_{tH}(\tau)$,  which allows to probe also
the chiral/flavor 
structure of the instanton-induced vertex. We suggest that this
quantity should be measured on the lattice. 

We studied the contribution of quark loops to chirality flipping dynamics,
by comparing the results of quenched and full simulations. 
We observed  dramatic quenching effects that can be  
naturally explained by instantons 
and suggests an interesting link between unitarity and topology.

\end{document}